# Single-Cavity Bi-Color Laser Enabled by Optical Anti-Parity-Time Symmetry

Yao Duan, Xingwang Zhang, Yimin Ding, and Xingjie Ni[*]

*Department of Electrical Engineering, The Pennsylvania State University, University Park, Pennsylvania 16802*

[*]*xingjie@psu.edu*

## Abstract

The exploration of quantum-inspired symmetries in optical systems has spawned promising physics and provided fertile ground for developing devices exhibiting exotic function- alities. Founded on the anti-parity-time (APT) symmetry that is enabled by both spatial and temporal interplay between gain and loss, we demonstrate theoretically and numerically bi-color lasing in a single micro-ring resonator with spatiotemporal modulation along its azimuthal direction. In contrast to conventional multi-mode lasers that have mixed-frequency output, our laser exhibits stable, demultiplexed, tunable bi-color emission at different output ports. Our APT-symmetry-based laser may point out a new route for realizing compact on-chip coherent multi-color light sources.

### 1. Introduction

Although mode competition in laser systems is usually considered as an obstacle for achieving stable single-mode operation, yet a multi-color laser supporting more than one mode that has stable, deterministic, frequency-separated coherent light emission is desirable owing to its wide applications in wavelength-division multiplexing [1], heterodyne interferometry [2], and full-color display [3]. Currently, there are two major ways to realize multi-color laser emission: combination of multiple single-mode laser cavities and multi-mode emission from a single cavity. Combining the lasers of all desired output wavelengths, e.g., individual red, green, and blue lasers to build a laser with white light output, is the most intuitive way to customize the color output [4–6]. But the convenience comes with the cost of low integration level owing to the requirement of additional optical interconnection modules. Multi-frequency emission from single cavity can be realized by excitation of multiple modes or stimulated emission processes [7–9]. Although those systems can be scaled down for dense integration, they suffer from instability of output power distribution among the desired modes due to variance of pumping power. Moreover, it is also difficult to separate different frequency components into different output channels without assistance from additional optical elements. Therefore, both of the approaches are not competent for creating compact on-chip multi-color laser sources. Previous works have tried to address these challenges by integrating subwavelength-spaced structures, e.g., metasurfaces, into gain medium to control the resonances. With proper designs, only the targeted wavelengths are enabled through superlattice plasmonic resonances [10, 11], which improve the stability as well as spatially manipulate the transverse wavevectors of different laser emissions. However, such difference in wavevectors is very small down to several hundreds of $\mu m^{-1}$ in free space, which is still difficult to spatially separate them without external modules. Therefore, a miniaturized, stable multi-color laser with demultiplexed lasing output is still long-sought-after.

Recently, the exploration of non-Hermitian quantum symmetries, especially parity-time (PT) symmetry in photonic systems has enabled new features and functionalities in laser systems [12,13]. Compared with other approaches that mainly rely on varying refractive indices to tune cavity resonances, the non-Hermitian symmetry-based optical systems are based on the modulation of gain and loss, i.e., the part of refractive index to manipulate the imaginary part of resonances. The peculiarity of PT-symmetry-induced properties have led to the development of new laser systems, such as loss induced revival of lasing [14], alleviation of mode competition [15] and lasing wavefront shaping [16, 17]. However, current reported PT symmetric lasers only distribute gain and loss in the spatial dimension. This spatial degree of freedom has limited control over the coupling among modes with the same frequency. It does not have the capability to control the mode coupling in the frequency domain, which is essential for achieving controllable multi-color lasing in a single cavity.

Here, we propose a new scheme for creating a single-cavity multi-color laser by engineering gain and loss in both spatial and temporal domains based on the concept of anti-parity-time (APT) symmetry. As a demonstration, we designed and numerically simulated a micro-ring bi-color laser with realistic physical parameters. The stabilized and tunable bi-color emission is protected by the modulation wavevector and modulation frequency. Moreover, as the laser emission is originated from two counter-propagating modes in the micro-ring cavity, the two frequency outputs are well-separated and can be used independently. We believe that this APT-symmetry-enabled bi-color lasing scheme provides a brand new route towards compact on-chip multi-color lasers, which could be promising coherent sources for future photonic integrated chips.

## 2. Theoretical Background and Working Principle

The theoretical foundation of our proposed bi-color laser lies on the optical APT symmetry. In contrast to PT symmetric Hamiltonians which fulfill the commutation relation $[\hat{H}, \hat{P}\hat{T}] = 0$ under combination of parity (P) and time-reversal (T) operations, APT symmetric Hamiltonians satisfy anti-commutation relation $\{\hat{H}, \hat{P}\hat{T}\} = 0$[18]. An APT symmetric Hamiltonian can be obtained from a PT symmetric one by multiplying the unit imaginary number $i$, $\hat{H}^{(APT)} = i\hat{H}^{(PT)}$. Such close connection between APT and PT symmetric systems has attracted great interest in investigating the physics of APT symmetry in various configurations in atomic [19], thermal [20], electrical [21] and optical [22–24] systems. However, how to exploit this new quantum symmetry into practical laser applications is still yet to be discussed.

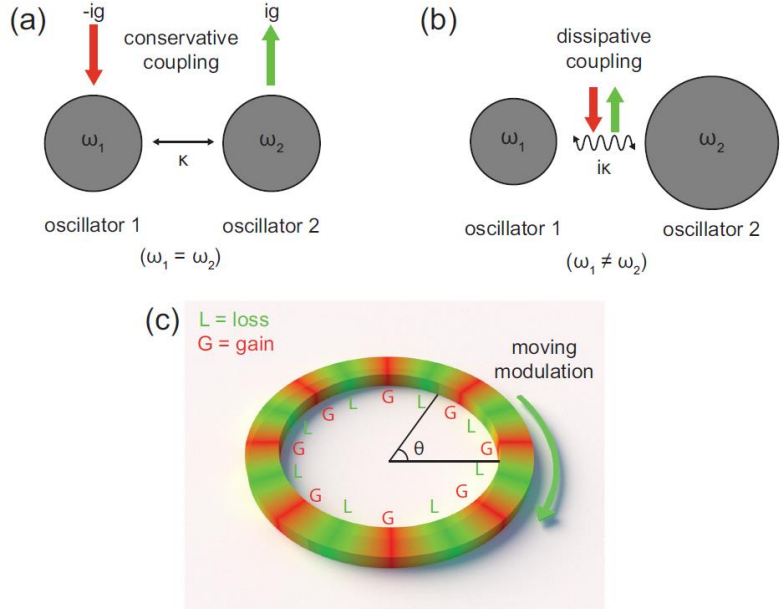

Fig. 1. Schematic illustration of the (**a**) PT and (**b**) APT symmetries in a two-oscillator system. *ig* (*ig*) and *K* represents gain (loss) and coupling coefficients, respectively. In the PT symmetric system, resonance frequencies $\omega_1$, $\omega_2$ of the two oscillators are identical, while they are different in the APT symmetric one. (**c**) The configuration of APT symmetric micro-ring resonator under spatiotemporal modulation for bi-color lasing. The dynamic imaginary grating of permittivity with alternating gain (G) and loss (L) is moving azimuthally along the ring. The moving direction is indicated by the green arrow.

Based on the relationship between PT and APT symmetries, we derived how the bi-color lasing mode can be realized in an APT symmetric system. Typically, a PT symmetric system can be described by two oscillators with an identical resonance frequency, yet opposite damping factors. They are coupled through conservative coupling process such as near field coupling (Fig. 1(a)). In contrast, an APT system can be achieved by two oscillators with a same damping factor but different resonance frequencies. Those oscillators are coupled through purely dissipative channels such as far field coupling, as described in Fig. 1(b), satisfying $\hat{H}^{(APT)} = i\hat{H}^{(PT)}$. We leverage two whispering gallery modes (WGMs) in a micro-ring resonator to realize an APT system. In order to obtain dissipative coupling component between two counter-propagating clockwise (CW) and counterclockwise (CCW) WGMs, we apply a moving grating with sinusoidally modulated imaginary part of permittivity azimuthally along the micro-ring resonator (Fig. 1(c)). The permittivity change can be expressed as $\Delta\varepsilon(\phi,t) = i\Delta\varepsilon_I \cos(L_m\phi - \omega_m t)$, where $\Delta\varepsilon_I$ is modulation depth, $L_m$ is azimuthal modulation wavevector, $\omega_m$ is modulation frequency, $\phi$ and $t$ are azimuthal angle and time respectively. This dynamic imaginary grating can be achieved through pumping the micro-ring resonator using the interference of two optical beams with slight different center frequencies [25].

To show that the system is APT symmetric, we analyzed a pair of degenerated CW and CCW WGMs in the micro-ring cavity with sinusoidal modulation profile. Assuming coupling only exists between these two modes with phase matching condition, the eigenmodes of this system can be obtained by solving the temporal coupled-mode equations below [26]

$$\hat{H}(t)\psi(t) = i\frac{d\psi(t)}{dt} \tag{1}$$

where the time-dependent state vector $\Psi(t)$ and Hamiltonian of the system $\hat{H}(t)$ are written as

$$\psi(t) = \begin{pmatrix} a_1 \\ a_2 \end{pmatrix}, \quad \hat{H}(t) = \begin{pmatrix} \omega_0 & i\kappa_m e^{-i\omega_m t} \\ i\kappa_m e^{i\omega_m t} & \omega_0 \end{pmatrix} \tag{2}$$

Here, $a_1$ and $a_2$ are the amplitudes of the CW and CCW modes, respectively, $\omega_0$ is the original resonance frequency of the two WGMs and $i\kappa_m e^{\pm i\omega_m t}$ is the time-varying coupling coefficient where $\kappa_m$ is proportional to the modulation depth of $\Delta\varepsilon_I$. It should be emphasized that $\omega_m$ can be either positive or negative in the Hamiltonian to indicate the modulation travels along CCW or CW direction respectively. Taking the gauge transformation $a_{1,2} = A_{1,2} e^{-i(\omega_0 \pm \omega_m/2)t}$ to remove the temporal variance, we get a time-independent $\hat{H}$

$$\hat{H} = \begin{pmatrix} -\omega_m/2 & i\kappa_m \\ i\kappa_m & \omega_m/2 \end{pmatrix} \tag{3}$$

We can see that the degeneracy between CW and CCW modes is broken in the new frame. The eigenvalues $\omega_{1,2}$ and the corresponding eigenvectors $\psi_{1,2}$ can be obtained

$$\omega_{1,2} = \pm\sqrt{\frac{\omega_m^2}{4} - \kappa_m^2} \tag{4}$$

$$\psi_1 = \begin{pmatrix} 1 \\ -i\Delta\omega/\kappa_m \end{pmatrix}, \quad \psi_2 = \begin{pmatrix} 1 \\ -i\kappa_m/\Delta\omega \end{pmatrix} \tag{5}$$

where $\Delta\omega = \sqrt{(\omega_m^2/4 - \kappa_m^2)} + \omega_m/2$. We note that the equation is solved under a gauge transformation $e^{-i(\omega_0 \mp \omega_m/2)t}$, which means the solutions in Eq. (4) are eigenfrequencies in the moving frames of which the rotating frequencies are $\mp\omega_m/2$. After changing back to the laboratory frame, the actual number of observable eigenfrequencies should be four in total, which are $\omega_{1,2} + \omega_m/2$ and $\omega_{1,2} - \omega_m/2$.

Looking back into the time-independent Hamiltonian, it is evident that $\{\hat{H}, \hat{P}\hat{T}\} = 0$. Here the parity operator $\hat{P}$ represents the mirror reflection given by the first Pauli matrix $\sigma_x$, which exchanges the spatial positions of these two optical modes and the time-reversal operator $\hat{T}$ is given by the complex conjugation. Therefore, we can verify that this system satisfies APT symmetry.

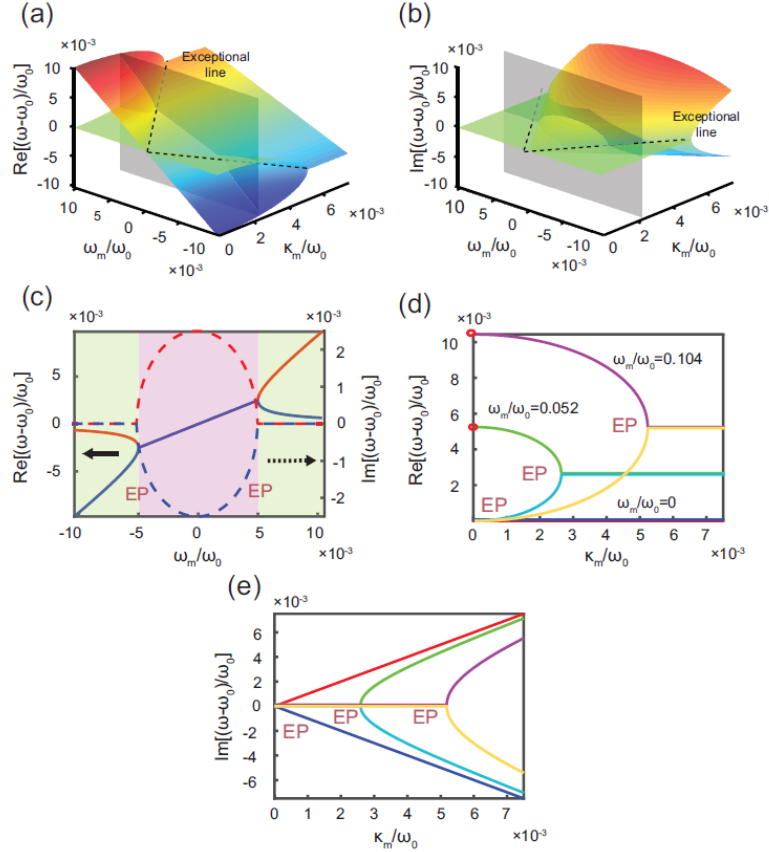

Fig. 2. (**a**) Real part and (**b**) imaginary part of normalized eigenfrequency spectra with respect to modulation frequency $\omega_m$ and coupling coefficient $\kappa_m$. The black dashed lines are the exceptional lines, where the two eigenstates degenerate. (**c**) The real (solid lines) and imaginary part (dashed lines) of eigenfrequencies under a fixed $\kappa_m$ indicated by the gray planes in (a) and (b), respectively. The yellow and red regions indicate broken APT and unbroken APT phases, respectively. (**d**) Real part and (**e**) imaginary part of the eigenfrequencies versus $\kappa_m$ with different $\omega_m$ (indicated by different colors). The red circles in (d) indicate the uncoupled cases ($\kappa_m = 0$) and there are no sidebands. The red and blue curves are the modes in a PT symmetric case as reported in [15] for reference.

The dependence of real and imaginary parts of normalized $\omega_{1,2} + \omega_m/2$ with different coupling coefficient $\kappa_m$ and temporal modulation frequency $\omega_m$ are shown in Fig. 2(a)(b), respectively. From both figures, the eigenfrequencies can be clearly classified into two regimes according to the existence of degeneracy. Taking a slice across both regimes (Fig. 2(c)), we can see that in the red regime, the real part is degenerate with a non-vanishing opposite imaginary part. In contrast, in the yellow regime the real part splits with a zero imaginary part. The boundary of two regimes is along the line where $\kappa_m = \pm\omega_m/2$ (the dashed lines in Fig. 2(a)(b)). Similar to those in PT symmetric systems, the eigenfrequency spectra in APT symmetric systems can transit from real to complex values. The transition boundaries are also called exceptional points (EPs) or exceptional lines (ELs). However, the properties of the corresponding eigenvectors are different between PT and APT symmetric cases. In a PT symmetric case, $\hat{P}\hat{T}\psi_{1,2} = \psi_{2,1}$ is satisfied in the regime where the eigenfrequency is complex. It is referred as broken PT phase, as $\hat{P}\hat{T}$ operates on $\psi_i$ do not lead to the same $\psi_i$ ($i$ = 1, 2). $\hat{P}\hat{T}\psi_{1,2} = \psi_{1,2}$ can be observed in the regime where eigenfrequency is purely real, which is referred as unbroken PT phase. This behavior is opposite in our APT symmetric case: $\hat{P}\hat{T}\psi_{1,2} = \psi_{1,2}$ is satisfied in the regime where the eigenfrequency is complex (unbroken APT phase), while $\hat{P}\hat{T}\psi_{1,2} = \psi_{2,1}$ is fulfilled when the eigenfrequency is purely real (broken APT phase). In other words, both PT and APT symmetries have broken and unbroken phases, but the spectral properties of each phase are completely different. Moreover, $\omega_m$ does not merely define the boundary between the broken and unbroken APT phases in the solution of $\omega_{1,2}$. It also contributes to the Doppler shift term $\omega_m/2$ which determines the frequencies of modes in the lab frame (with respect to the moving frames under the gauge transformation). Compared with the PT symmetry with only limited control over the phase transition, our APT system has additional degrees of freedom in manipulating spectral mode frequencies.

This spatiotemporal modulation enabled APT system with modes of opposite imaginary parts in the unbroken APT regime provides the foundation for creating our bi-color laser. In Fig. 2(c), the blue curve in the red region represents the mode of which the eigenfrequency has a negative imaginary part, indicating persistent amplification of the energy in this mode. With a sufficient large modulation depth $\kappa_m > |\omega_m|/2$, lasing from this mode is possible where the EP works as the lasing threshold since the mode begins to amplify beyond it towards the unbroken APT regime. Considering the frequency shifting term, two lasing modes with eigenfrequencies differed by $\omega_m$ can be observed in the system. Furthermore, if we assume the non-Hermitian coupling coefficient $\kappa_m$ is much larger than $\omega_m$, the corresponding eigenvector of both amplifying modes will be $\psi_1 = [1, 1]^T$ approximately. It indicates that the two lasing modes are actually the CW and CCW modes with broken degeneracy, respectively.

While we can only get a pair of degenerate CW and CCW lasing modes in a typical PT symmetric scheme since the real part of their eigenfrequencies are identical, the dynamic modulation in our APT symmetric scheme enables frequency detuning of $\pm\omega_m/2$ for those two lasing modes, respectively, in the unbroken APT phase. This property makes possible bi-color lasing in a single micro-ring cavity. Most importantly, those lasing modes propagate in opposite directions hence are easy to separate spatially which is not possible for conventional multi-mode lasers.

Considering $\kappa_m$ and $\omega_m$ are separately determined by the spatiotemporal modulation depth and frequency, respectively, this APT system offers flexible control over the lasing properties. For example, the same phase transition process of $\omega_{1,2} + \omega_m/2$ can also be observed with determined $\omega_m$ and varying $\kappa_m$ (Fig. 2(d)(e)). The EP position is shifted towards larger $\kappa_m$ with larger $\omega_m$. As the bi-color laser works only in the broken APT phase, the position of the EP can be effectively treated as an on/off switch for the bi-color lasing operation. Since the EP is located at $\kappa_m = \pm\omega_m/2$, both $\omega_m$ and $\kappa_m$ determine the lasing threshold. Moreover, $\omega_m$ also tune frequency difference between two lasing modes.

## 3. Micro-Ring Bi-Color Design and Numerical Validation

The unique features of our APT symmetry in the unbroken APT phase provide the theoretical keystone for realizing a demultiplexed multi-color laser. We chose realistic material parameters to construct our bi-color laser. Our micro-ring resonator consists of a semiconductor gain material, InGaAsP multi quantum well (MQW), which sits on InP substrate. We kept the same sinusoidal profile of modulation as discussed in Section 2. The WGMs of different orders supported by the mirco-ring cavity are denoted by $|\pm l\rangle$, where the positive and negative signs indicate CW and

CCW modes, respectively. The mode number $l$ is an integer that can be calculated from $l = Dn_{\text{eff}}/\lambda$, where $D$ is the perimeter of the micro-ring, $n_{\text{eff}}$ is the effective index of the WGM, and $\lambda$ is the free-space wavelength. We choose $|\pm l\rangle$ as two uncoupled modes and introduce the coupling through spatial phase-matching condition $L_m = 2l$. Since $\Delta\varepsilon_I$ is proportional to $\kappa_m$ between the $|\pm l\rangle$ modes (see Appendix E for detailed derivation), for the sake of simplicity we use directly $\Delta\varepsilon_I$ instead of $\kappa_m$ in the following discussions.

We established a numerical model and conducted full-wave simulations of the proposed bi-color micro-ring laser using the finite different time-domain (FDTD) method. The imaginary part of the permittivity of the micro-ring is spatiotemporally modulated in the azimuthal direction (See Appendix A for details of the model). In order to extract the lasing output, we used two straight waveguides evanescently coupled to the micro-ring in an add-drop configuration. With $l = 21$, $L_m = 42$, and spatiotemporal modulation frequency $\omega_m = 1$ THz, the evolution of the resonances at different modulation depth $\Delta\varepsilon_I$ can be observed in the normalized transmission and lasing spectra (Fig. 3(a)(b)). The definitions of the ports are depicted in the inset of Fig. 3. While without modulation the signal enters the structure through the left-side waveguide from port 1 and couples to the CCW resonance, creating a dip in the transmission spectrum $T_{12}$ and a peak in the reflection spectrum $T_{11}$ around 1.555 μm.

Increasing the modulation depth of permittivity, total four resonances are clearly observed (Fig. 3(a)(b)). They can be classified into two categories in which the resonance frequencies differ by 1 THz (or 8 nm in term of wavelength). In each category, the frequency difference between two resonances persistently reduces while increasing $\Delta\varepsilon_I$ and finally becomes zero when $\Delta\varepsilon_I = 0.415$. This indicates the phase transition across this point. In the red region, two amplified peaks are observed in the normalized spectra. We would like to note that there is no signal input in this region. The numerically acquired resonance frequencies at different $\Delta\varepsilon_I$ agree well with our prediction based on the APT symmetry theory across both yellow and red regions (Fig. 3(c)).

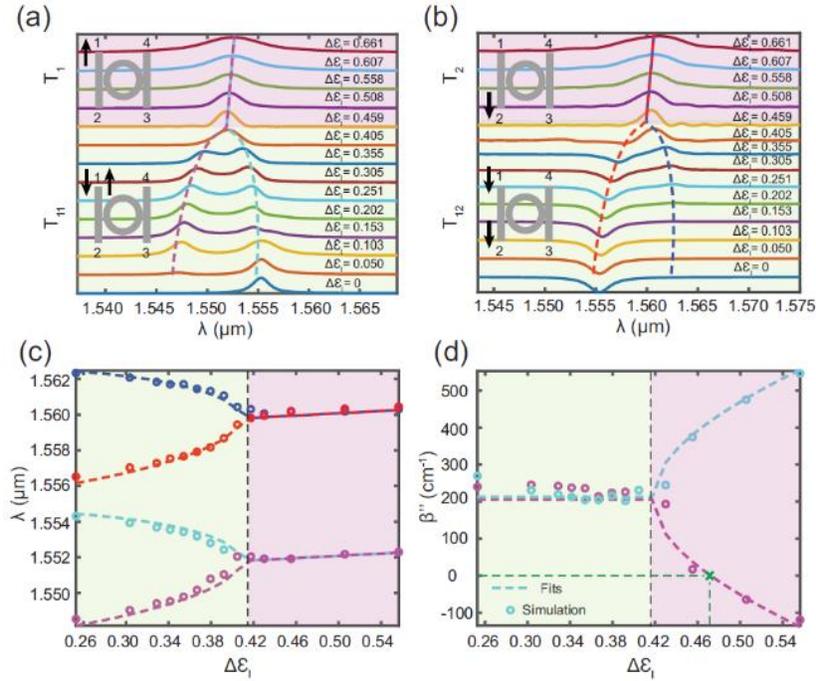

Fig. 3. (**a**) Normalized reflection spectra $T_{11}$ and lasing spectra $T_1$, (**b**) normalized transmission spectra $T_{12}$ and lasing spectra $T_2$ under different modulation depth $\Delta\varepsilon_I$ with a fixed modulation frequency $\omega_m = 1$ THz (See Fig. 6 for $T_{13}$ and $T_{14}$). The definitions of the ports, $T_{11}$, $T_{12}$, $T_1$, and $T_2$ are depicted in the insets. The yellow and red regimes indicate the broken and unbroken APT phases, respectively. (c) The real parts of the resonance frequencies and (d) the imaginary parts of resonance wavevectors extracted from the full-wave simulations (circles) and theoretical calculations based on the APT symmetry (dashed lines). The colors of the dashed curves in (c)-(d) correspond to the lines of the same color in (a)-(b).

To further verify whether this change in the spectra is indeed the phase transition between broken and unbroken APT phases, we obtained the imaginary parts of resonance wavevectors by fitting each spectrum with two Lorentz functions. The imaginary parts of the wavevectors $\beta''$ then can be acquired through $\beta'' = \Delta\omega''/c = 2n\pi\Delta\lambda/\lambda$, where $c, n, \lambda, \Delta\lambda$ are speed of light, the refractive index of material, center wavelength, and the linewidth of the resonance, respectively. The obtained values from simulations again match well with those calculated from the APT theory (Fig. 3(d)). We can see that bifurcation occurs between two resonances in the imaginary spectrum in the red region while they coalesce in the yellow region, which also matches well with our theoretical prediction, indicating the phase transition of the APT symmetry. Additionally, in our simulation, we noticed that $\beta''$ is not zero at the EP. This owes to the inherent loss in the system (Fig. 3(d)).

In the unbroken APT phase, two spectral lines separated by $\omega_m$ can be detected respectively in all four ports. Lasing output at $\lambda = 1.560$ µm is observed at port 2 and 4, which results from the CW mode; Meanwhile, $\lambda = 1.552$ µm laser line is observed at port 1 and 3, which comes from the CCW mode (Fig. 4(a)(b)). Therefore, it is verified that two laser lines with different wavelengths can be generated simultaneously in a single micro-ring cavity and well separated inherently due to the directionality of the WGM modes. The bi-color lasing frequencies can be continuously tuned through varying the modulation frequency $\omega_m$ within $|\omega_m| < 2\kappa_m$ (Fig. 4(c)). In addition, the frequency tuning range for the bi-color lasing mode can be extended by increasing $\Delta\varepsilon_I$.

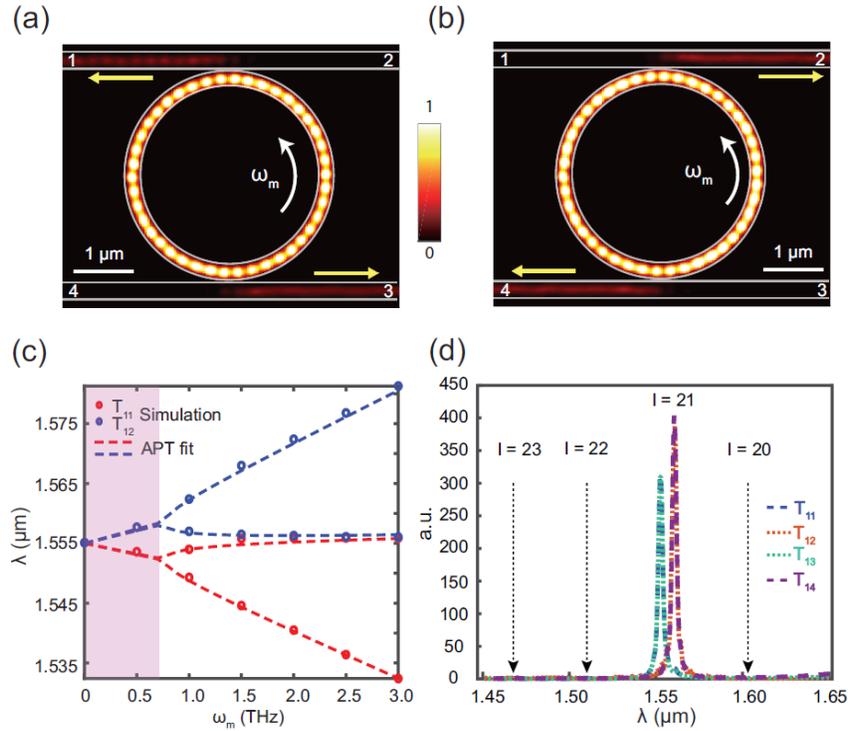

Fig. 4. The normalized intensity distribution of the micro-ring system at (a) $\lambda = 1.552$ µm and (b) $\lambda = 1.560$ µm with $\Delta\varepsilon_I = 0.508$ and $\omega_m = 1$ THz. The white

lines outline the geometry of the micro-ring as well as the coupling waveguides. The white arrows indicate the traveling direction of spatiotemporal modulation and the yellow arrows show the output lasing direction. The ports are numbered in the same way as those shown in Fig. 3. (**c**) The real parts of resonance frequencies extracted from full-wave simulations (circles) and the calculated ones based on our APT theory (dashed lines) with a fixed $\Delta\varepsilon_l$ = 0.315 while varying $\omega_m$. The red regime indicates the unbroken APT phase. (**d**) Lasing spectra from all ports with same modulation depth as in (a) and (b). $l$ is the azimuthal order of the WGM.

Moreover, our APT-based laser significantly reduces unnecessary mode competition. In our micro-ring laser, only the desired WGMs $|\pm 21\rangle$ can be coupled to each other through dynamic modulation under phase matching condition ($L_m = 42$). The modulation frequency is smaller than free spectral range, hence no coupling among neighboring WGMs were introduced. In addition, the dynamic modulation re-distribute the optical gain such that it maximizes amplification for the desired mode while suppresses all other WGMs (Appendix F). Therefore, lasing from only the desired modes ($|\pm 21\rangle$) can be achieved (Fig. 4(d)). This reduction of mode competition improves the stability of bi-color lasing under various pumping condition.

In addition to the laser operation we demonstrated in unbroken APT phase, nonreciprocal light propagation can be observed in the broken APT phase of our system. The temporal modulation of permittivity shifts the resonances oppositely for CW and CCW modes and breaks the reciprocity of the system. This can be verified by comparing the transmission spectra $T_{12}$ and $T_{21}$ at different input port respectively (Appendix E). This nonreciprocal behavior in the broken APT regime could lead to realization of other useful optical elements such as an optical isolator.

## 4. Discussion on Experimental Realization

In practice, the azimuthal spatiotemporal modulation can be realized by interference between two laser pump beams of different frequencies carrying different orbital angular momenta (OAM). The imprinted vortex phase front of different OAM orders will provide required $L_m$. In the past, we have achieved spatiotemporal modulation of the real part of permittivity along a straight line using a similar technique, where we used two laser pump beams with slightly different frequencies to generate a traveling wave interference pattern [25, 27]. The temporal modulation frequency can easily reach up to several terahertz by exploiting the instantaneous third-order nonlinear response of material [28]. However, the maximum modulation speed of gain is limited by the relaxation time of the excited carriers in the material which is typically around nanosecond scale in semiconductors. Hence, the largest modulation frequency achievable for gain is in the GHz range. In order to have higher modulation frequency, a feasible solution is to apply a spatiotemporal loss modulation instead of gain modulation with materials of ultrafast response. For example, the relaxation time of carriers in graphene can be as short as several hundreds of femtoseconds [29], and tuning the loss of graphene under optical pumping has already been utilized for ultrafast optical modulators [30]. Therefore, we envision that by placing thin layers of graphene on top of our optical structures, we can effectively modulate the imaginary part of the permittivity at a high frequency.

## 5. Conclusion

In conclusion, we theoretically proposed and numerically validated an APT symmetric bi-color laser enabled by spatiotemporal modulation of gain/loss distribution in a micro-ring resonator. Tunable modulation frequency and modulation depth provide two degrees of freedom for real-time manipulation of the lasing dynamics. This APT-symmetry-enabled bi-color lasing scheme can be generalized to the multi-color cases with more than two lasing modes. Exploiting the non-Hermitian symmetries in optical system, we can expect the revolution of current lasers in terms of functionality, stability, size, etc., which could enable a plethora of applications such as on-chip coherent light sources for communications, remote sensing, and displays.